\documentstyle[12pt,aaspp4]{article}

\lefthead{Elmegreen, Elmegreen, Chromey, Fine}
\righthead{Dust Streamers in M86 from VCC 882}
\slugcomment{{\it Astronomical Journal}, August 2000, in press}
\begin{document} 

\title{Dust Streamers in the Virgo Galaxy M86 from Ram
Pressure Stripping of its Companion VCC 882} 

\author{Debra Meloy Elmegreen\altaffilmark{1}, Bruce G.
Elmegreen\altaffilmark{2}, 
Frederick R. Chromey\altaffilmark{1}, 
Michael S. Fine\altaffilmark{1,3}}

\altaffiltext{1}{Department of Physics and Astronomy, Vassar College,
Poughkeepsie, NY 12604; e--mail: elmegreen@vassar.edu, chromey@vassar.edu}
\altaffiltext{2}{IBM Research Division, T.J. Watson Research Center, P.O. Box
218, Yorktown Heights, NY 10598; e--mail: bge@watson.ibm.com}
\altaffiltext{3}{Department of Physics and Astronomy, Colgate University,
Hamilton, NY, 13346 e--mail: msfine@mail.colgate.edu}

\begin{abstract}
The giant elliptical galaxy M86 in Virgo has a $\sim28$ kpc long dust
trail inside its optical halo that points toward the nucleated dwarf
elliptical galaxy, VCC 882.  The trail seems to be stripped material
from the dwarf.  Extinction measurements suggest that the ratio of the
total gas mass in the trail to the blue luminosity of the dwarf is about
unity, which is comparable to such ratios in dwarf irregular galaxies.
The ram pressure experienced by the dwarf galaxy in the hot gaseous
halo of M86 was comparable to the internal gravitational binding energy
density of the presumed former gas disk in VCC 882. Published numerical
models of this case are consistent with the overall trail-like morphology
observed here.  Three concentrations in the trail may be evidence for
the predicted periodicity of the mass loss.  The evaporation time of
the trail is comparable to the trail age obtained from the relative
speed of the galaxies and the trail length.  Thus the trail could be
continuously formed from stripped replenished gas if the VCC 882 orbit is
bound. However, the high gas mass and the low expected replenishment rate
suggest that this is only the first stripping event.  Implications for
the origin of nucleated dwarf ellipticals are briefly discussed.

\end{abstract}

\section{Introduction}

M86 (NGC 4406) is a bright elliptical (E3/S0) galaxy located in the
Virgo Cluster at a distance of $\sim18.3$ Mpc (Capaccioli et al. 1990).
It has a redshift of $-227$ km s$^{-1}$ (Binggeli et al. 1985), while the
mean heliocentric velocity of the cluster as a whole is 1050 km s$^{-1}$
(Binggeli et al. 1993). M86 is thought to be on an orbit passing through
the core of the cluster approximately every 5 billion years (Forman et
al. 1979). It is an X-ray object and appears to be a weak radio
source (Laing et al. 1983; Fabbiano et al. 1992; Rangarajan et al.
1995). A plume of X-ray, HI, and infrared emission from M86 suggests
that its interstellar medium was swept back by the ram pressure from its
motion through the intracluster medium (Forman et al. 1979; Fabian,
Schwarz, \& Forman 1980; Takeda, Nulsen, \& Fabian 1984; Bregman \&
Roberts 1990; Knapp et al. 1989; White et al. 1991). There is also an
optical asymmetry to M86 that gives it a slightly enhanced emission
along the plume (Nulsen \& Carter 1987).

Here we discuss a nucleated dwarf elliptical galaxy, VCC 882 (NGC
4406B; Binggeli et al. 1985), that lies just to the northeast of M86,
inside its projected stellar halo. Deep CCD images show a 28 kpc long
dust trail inside M86 that appears to follow VCC 882 in its orbit. This
trail is possibly the result of ram pressure stripping of gas originally
inside VCC 882 that was removed by the high pressure of its motion
through the hot gaseous halo of M86. 
The gas mass obtained from the extinction in the trail is 
consistent with this former connection to VCC 882. Other evidence for an
interaction between M86 and VCC 882 could be an isophotal twist in the
central 8 to 80 arcsec of M86 (Bender \& Mollenhoff 1987), and the
asymmetric outer isophotes of M86 (Nulsen \& Carter 1987). 

Gas stripping is pervasive within this core region of the Virgo cluster.
Three spirals are close to the M86/VCC 882 pair in projection: NGC 4438,
NGC 4388, and NGC 4402. There is also a nearby dwarf galaxy, IC 3355.
The spirals have been severely stripped of their outer HI disks
(Warmels 1986; Hoffman, Helou, \& Salpeter, 1988;
Cayatte et al. 1990), but the dwarf looks normal. The
peculiar negative velocity of M86 is also shared by NGC 4402, NGC 4438,
and IC 3355, as if at least some of these galaxies are comoving like a
group through the Virgo cluster (Kotanyi \& Ekers 1983). Detailed
studies of one of these galaxies, NGC 4438, suggest that ram pressure
from its motion through the Virgo intracluster medium has visibly
distorted its outer disk (Arp 1966), pushing the gas and star formation
off the normal plane, and producing a short diffuse trail of radio
continuum and X-ray emission (Kotanyi \& Ekers 1983; Kotanyi, van Gorkom
\& Ekers 1983). 
A second galaxy in this group, NGC 4388, has been
studied by Pogge (1988), Petitjean \& Durret (1993) and Veilleux et al.
(1999), with mixed results on a stripped origin for extraplanar
material.   The spiral
galaxy NGC 4569, at an equal distance from M87 but on the eastern side,
has a negative velocity too, along with an anemic classification (van
den Bergh 1976) and depletion in HI (Cayatte et al. 1990). 
Another dwarf galaxy, IC 3475, is closer in projection to M87
than IC 3355 and is highly stripped of HI (Vigroux et al. 1986), while
another anemic spiral, NGC 4548, is about twice as far from M87 as these
others and has a distorted outer HI disk (Vollmer et al. 1999).

Ram pressure seems to have stripped the Virgo spiral NGC 4694 also. This
galaxy is far to the east of the core region in Virgo. It has a 36 kpc
long trail of HI streaming off to the west, with a linear velocity 
gradient along
the trail that smoothly connects it with the galaxy (van Driel
\& van Woerden 1989). A very faint dwarf galaxy is also in the trail,
where the HI column density peaks. 

All of these cases suggest stripping from the motion of a galaxy through
the hot, low-density, intracluster medium inside Virgo. The case
discussed here differs because VCC 882 was apparently stripped by its
motion through the much denser hot gas associated with the elliptical galaxy
M86. Its relative speed is just as large as in the other cases,
exceeding 1000 km s$^{-1}$, but the ram pressure it felt must have been
much larger because of the higher ambient density. 

A similar case is the
Virgo dwarf galaxy UGC 7636, which was apparently stripped as it moved
through the giant elliptical, NGC 4472. The evidence for this is:
a gas cloud to the side of UGC 7636 (Sancisi, Thonnard \& Ekers 1987)
that
has the right mass to have been formerly part of UGC 7636 (Patterson \&
Thuan 1992; Irwin \& Sarazin 1996); a 30 kpc long optical trail
of luminous debris adjacent to the cloud and the dwarf (McNamara et al.
1994); absorption of the elliptical galaxy X-ray radiation by the HI
cloud (Irwin \& Sarazin 1996), and an oxygen abundance in an HII region
of the cloud that is consistent with the abundance expected for UGC 7636
(Lee et al. 2000). 

Previous photometric studies of Virgo cluster galaxies were made by
Binggeli et al. (1985, 1993), Bender \& Mollenhoff (1987), and Caon et
al. (1994). Katsiyannis et al. (1998) co-added thirteen Schmidt
exposures to produce a deep R-band image of the Virgo southeast region,
from which they studied extended and overlapping halos.  The dust
trail discussed here did not appear in these other images. 

\section{Observations and Data Reduction}

We obtained 92 images in B, V, and I over an approximately 1 square
degree field in the southeast center of the Virgo Cluster using the
Burrell Schmidt 0.6-m telescope at Kitt Peak National Observatory on
17-22 March 1999. The plate scale is 1.6 arcsec per pixel, or about 142
pc per pixel at the assumed distance of 18.3 Mpc. The images were
reduced using standard IRAF procedures. Flat fields were made from
global sky flats taken over the entire observing run. The images were
mosaicked to produce combined images with total exposure times of
approximately 6.5 hours in each band. 

Figures 1a and 1b show a mosaic of the central square degree of the
images in B band. The left-hand image is low-contrast, showing the giant
ellipticals M84 and M86 as well as the small companion VCC 882, which is
normally lost in the bright light of M86. The
right-hand image is shown with high contrast to emphasize the outer
extents of the large galaxies.

Enhanced images of M86 are in Figure 2. The top left frame shows the
logarithm of the intensity in B band with the angular scale indicated.
Dark dust features are labeled. Feature A is 2.05 arcmin east
of center, and Feature B is 3.7 arcmin southeast of center, midway
between two bright foreground stars. These features are more prominent
in the other enhanced images. In the top right, which is a (B-I) color
map, Feature A has a fork at its mid-point and a faint extension, marked
C, towards VCC 882. The lower left frame is an unsharp-masked image,
made by smoothing the original with a Gaussian filter 5 pixels wide and
subtracting this from the original. Most of the stars have been removed
from this image by replacing the corresponding pixels with the average
surrounding background. A star at the top of the eastern dust feature
remains, as well as another due west of the center of M86 and one near
VCC 882. The lower right-hand frame is an "embossed" B-band image made
in Adobe Photoshop; the three-dimensional perspective is the result of
an apparent "illumination" from the west. Here, the dust features are
black against the gray background. 

Previous published images of M86 did not show the dust features present
in Figure 2. They are too faint to appear on high-contrast prints.
Bregman \& Roberts (1990) mapped this region in HI and reported a
private communication with J. L. Tonry, who noticed dust patches
$100^{\prime\prime}$ east of center. These are probably what we see
here. The associated gas did not show up in H I emission because it was
below Bregman \& Roberts' detection limit. 

At the assumed distance of 18.3 Mpc, Features A and B lie at projected
distances of 10.9 kpc and 19.7 kpc from the center of M86. The latter is
comparable to the 19.5 kpc radius of M86 at a surface brightness of 25
mag arcsec$^{-1}$ (de Vaucouleurs et al. 1991). The outermost halo of
M86 shown in the right-hand image of Figure 1 has a much larger radius
of 62 kpc. Feature A is approximately 130$^{\prime\prime}$ ($=11.5$ kpc)
long from north to south, and its average width is 6.4$^{\prime\prime}$
($=570$ pc). Feature B is $\sim63^{\prime\prime}$ ($=5.5$ kpc) long
and $\sim19^{\prime\prime}$ ($=1.7$ kpc) wide, while Feature C is
$\sim27^{\prime\prime}$ ($=2.4$ kpc) long and $\sim4.8^{\prime\prime}$ 
($=430$ pc) wide. The total length of the trail is 28 kpc. 

In order to estimate the extinction of the dust features, we determined
the magnitude difference between them and their surrounding regions in
each passband. East-west intensity cuts were made at 28 points along the
length of the trail; 17 of these cuts that are
on Feature A are shown in Figure 3. The midpoint of the
cut was at the approximate position of the trail.
The dashed lines are averages of all 17 cuts. 
The magnitude differences between the dust and the surrounding
regions varied from 0 to 0.2 in B band and from 0 to 0.15 in V
band, with uncertainties of about 0.02 mag. The dust features do not
show up well in I band. 

Figure 4 shows the magnitude differences in the dust features for the B
and V bands as a function of position from south to north.  The ratio
of these magnitude differences, $\Delta m_B/\Delta m_V$, is shown at
the top.  Least squares fits to feature A are shown as
dotted lines. 
The endpoints of the curves are the southeastern dust Feature
B and the northernmost extension, Feature C, near VCC 882.  The magnitude
differences decrease from south to north in both B and I bands. The
ratio of the magnitude differences in B and V bands is slightly larger
than in a Whitford reddening law, where it would be 1.3.  The data are
not accurate enough to tell if this ratio changes along the length of
the trail. It should decrease with greater depth inside the halo of M86
because foreground stars wash out the color difference in the trail,
but such a decrease can be offset by changes in 
intrinsic extinction. 
The opacity and depth of the dust trail are
estimated in Section \ref{sect:opacity} from simple radiative transfer.

Figure 5 shows the B and I-band radial profiles along the major and
minor axes of VCC 882. The profiles were made after first removing the
underlying light from M86 by rotating the image $180^\circ$ and
subtracting it from the original. The profiles are well-fit by an
exponential, which is the case for many Virgo dwarf ellipticals (Vader
\& Chaboyer 1994; Ryden
et al. 1999). The scale length is 3.8$^{\prime\prime} = 340$ pc. The
(B-I) color profiles along the major and minor axes are shown at the
bottom of Figure 5 as solid and dashed lines. The southern side of VCC
882, nearest M86, is bluer than the northwest side, while the color
profile along the major axis is symmetric and flat.

The isophotal contours of VCC 882 are in Figure 6. They are
symmetric like the color profile, also showing no evidence of a tidal tail.
In contrast, the interaction of the dwarf galaxy UGC 7636 and the giant
elliptical NGC 4472 shows a blue tidal tail in the dwarf 
(Patterson \& Thuan 1992). An ellipse fit for 
VCC 882 gives a position angle of $116^\circ\pm4^\circ$ and an
ellipticity of $0.26\pm0.02$. The inner $2^{\prime\prime}$ region in Figure 6 
shows an isophotal twist. 

\section{Grain Size, Opacity and Geometric Depth}
\label{sect:opacity}

The magnitude difference between the dust feature and the surrounding
field can be converted into an opacity and a relative depth through the
M86 halo stars if the intrinsic ratio of $\Delta m_B/\Delta m_V$ is known.
Denoting this ratio by $C$ ($=4/3$ for a Whitford reddening law),
we can write the observed intensities in V and B bands in terms of
the intensities of the halo light behind and in front of the cloud,
$I_0$ and $I_1$, respectively, and in terms of the opacity in $V$, $\tau_V$:
\begin{equation} I_V^{cloud}=I_0e^{-\tau_V}+I_1 \;\;\;;\;\;\;
I_B^{cloud}=XI_0e^{-C\tau_V}+XI_1 \end{equation} where $X$ is the
ratio of B to V intensity to the side of the cloud: \begin{equation}
I_V^{side}=I_0+I_1 \;\;\;;\;\;\; I_B^{side}=XI_0+XI_1.\end{equation}
The magnitude differences may be written in the form
$I_V^{cloud}/I_V^{side}=10^{-0.4\Delta m_V}$.  Thus we can express the
opacity $\tau_V$ in terms of the magnitude differences: \begin{equation}
{{1-e^{-C\tau_V}}\over {1-e^{-\tau_V}}}=
{{1-10^{-0.4\Delta m_B}}\over {1-10^{-0.4\Delta m_V}}},
\end{equation}
and we can express the relative depth of the cloud in the M86 halo as
\begin{equation}
{{I_1}\over{I_0}}={{10^{-0.4\Delta m_V}-e^{-\tau_V}}
\over{1-10^{-0.4\Delta m_V}}}.
\end{equation}

Figure 7 shows the V-band opacity and the relative depth of feature A for
four values of $C$, using the linear fits to $\Delta m_V$ and $\Delta m_B$
that are in figure 4.  The opacity generally decreases to the north, and
the depth increases.  The $C$ values are all required to be
larger than the maximum of $\Delta m_B/\Delta m_V$, because this maximum
is the value $C$ would have for a foreground cloud. In the figure,
the value of $C$ is given in terms of this maximum, which is 2.36.
The solution is not reliable when $C$ is close to the maximum, as shown
by the sudden decrease of the depth in the northern part of the trail for
the $C=max(\Delta m_B/\Delta m_V)$ case.  Large values of $C$ suggest that
the dust grains are smaller in the trail than they are in the Milky Way.
The measurements are too inaccurate to be conclusive, however. 

\section{Discussion}

The dwarf galaxy VCC 882 appears to have lost much of
its interstellar medium during a recent passage through the X-ray
emitting gas of M86. The variation of extinction and depth 
along the trail of this
debris, and the positive relative velocity of VCC 882, suggest that the
orbit of the dwarf took it from the lower-left foreground of M86 to the
upper middle or background. Ram pressure stripping is the mostly likely cause for
the gas removal, rather than tidal stripping, because the stellar
distribution in the dwarf is hardly affected by the encounter, and
ram-pressure stripping affects only the gas.

The total mass of the dust features was estimated by assuming a standard
HI column density of $1.9\times10^{21}$ cm$^{-2}$ for A$_V$=1 mag
(Bohlin, Savage \& Drake 1978), and multiplying the resulting column
density by the total area of the features. For a magnitude difference of
A$_V$=0.1 (Fig. 4),
the above measurements give a total mass of
$3.4\times10^7$ M$_\odot$. If the dwarf is deficient in dust, as is
common for small galaxies with low metallicities, then the gas mass in
the trail will be larger. For a metallicity of [Fe/H]=$-0.59\pm0.42$
(Brodie \& Huchra 1991), a proportionately lower extinction-to-gas ratio
would increase the trail mass by a factor of $\sim4$ to
$\sim1.4\times10^8$ M$_\odot$. We shall use this mass below. 

If the radiative transfer model is correct, then the opacity in the trail
can be larger than 0.1 mag; Figure 7 suggests it might be as large as 1
mag. The gas mass would then increase in proportion.  We do not consider
this model to be accurate, however, because of measurement errors in the
magnitude differences and because the results are very sensitive to
the unknown ratio $C$.  More accurate measurements of the trail should
improve this situation.

The timescale for the stripping can be estimated from the trail length,
which is $\sim28$ kpc, and the relative speed of VCC 882 inside M86,
$v_{rel}=1328$ km s$^{-1}$ (the heliocentric velocity of VCC 882 and M86
are 1101 km s$^{-1}$ and -227 km s$^{-1}$, respectively). 
The ratio of these numbers,
15 My, is a measure of the timescale for the trail to be deposited,
aside from projection effects. 

There is no measure of the mass of VCC 882, but we estimate the
absolute magnitude in B band to be $\sim-16$ mag based on a comparison
of our counts with photometric sources in the field. For comparison,
observations by Binggeli, Sandage, \& Tammann (1985), Harris (1991),
and Cohen (1988) convert to absolute B magnitudes of $-14.6$, $-16.2$,
and $-15.2$, respectively, using a distance of 18.3 Mpc.  Our estimate
of $\sim-16$ mag makes the galaxy luminosity $L_{B}\sim3.8\times10^8$
L$_\odot$. If the gas in the trail was formerly part of the VCC~882
galaxy, then the ratio of the HI mass to the stellar luminosity was about
unity, which is comparable to that for dwarf irregular galaxies (Roberts
1969; Swatters 1999) but high for nucleated ellipticals. 
This makes the proposed VCC 882
predecessor resemble ESO 359-G29, which is a gas-rich
nucleated, dwarf elliptical (Sandage \& Fomalont 1993).
For a normal stellar population with $M/L_B\sim2$
M$_\odot$/L$_\odot$ (Sandage \& Fomalont 1993;
Hirashita, Takeuchi, \& Tamura 1998), the galaxy
mass would be $M_{gal}\sim8\times10^8$ M$_\odot$.  This, combined with
the exponential scale length of $R_{gal}=340$ pc, gives a characteristic
virial speed of  $GM_{gal}/(5R_{gal})^{1/2}=45$ km s$^{-1}$.

The presence of globular clusters around VCC 882 (Cohen 1988), as well
as the symmetric isophotal contours in the outer parts of this galaxy
(Fig. 6), suggest that ram pressure stripping of the gas may not have
been accompanied by significant tidal disruption of the stars. However,
tidal disruption detaches stars and globular clusters slowly from
the gravitational pull of VCC 882; it need not have moved the stars very
far yet from the galaxy center. If we multiply the 15 My interaction
time by the 45 km s$^{-1}$ virial speed inside VCC 882, we get a plausible
drift distance of $\sim 675$ kpc, to within a factor of$\sim3$
uncertainty from projection effects. This is not a large enough distance
for the globulars to have migrated significantly from the center of VCC
882, considering that globulars are usually seen in the outer parts of
galaxies anyway. Thus the dwarf could have been exposed to a significant
tidal force from M86, but we would not necessarily have noticed the effects
of this force yet. 

The ram pressure force on the gas can have a very different effect than
the tidal force on the stars. The average density in the X-ray halo of
M86 is $n\sim0.01$ cm$^{-3}$ (Thomas et al. 1987), and the VCC 882
orbital speed discussed above is $v_{rel}=1328$ km s$^{-1}$. Thus the
average ram pressure on the VCC 882 gas is $P=n\mu v_{rel}^2\sim
3\times10^{6}$ K cm$^{-3}$ for mean atomic weight
$\mu=2.2\times10^{-24}$ g. Such a pressure can strip the ISM from VCC
882 all at once if it exceeds the gravitational binding energy density
that
the gas would have had there. This energy density comes from the former
gas density multiplied by the square of the escape velocity in the
dwarf. This condition for stripping may be written \begin{equation}
\rho_{\rm external}v_{rel}^2>\rho_{\rm internal}v_{\rm escape}^2,
\label{eq:strip} \end{equation} which is comparable to the conditions
written by Gunn \& Gott (1972) for disk stripping and Takeda et al.
(1984) for spheroid stripping. We take the escape speed to be $2^{1/2}$
times the virial speed derived above. 

The {\it effective} density that the gas had when it was inside VCC 882,
if this is where it came from, can be estimated from the current gas
mass $M\sim1.4\times10^8$ M$_\odot$ in the trail and from the exponential
scale length $R_{gal}=340$ pc of the galaxy. This naively gives a
density of $3M/(4\pi R_{gal}^3\mu)\sim26$ cm$^{-3}$ for a spherical
distribution. This is high for an average ISM density, but it is not
supposed to be the real density, only the effective density for the
stripping calculation. With this density, the gravitational energy
density gives a self-binding pressure of $1.7\times10^7$ K cm$^{-3}$,
which is $\sim5\times$ larger than the ram pressure. 

We get about the same result if we consider a more realistic disk-like
structure for the gas before it was stripped. Then the expression for
self-binding pressure is $2\pi G \sigma_{tot}\sigma_{gas}$ (Gunn \& Gott
1972) for total and gaseous column densities $\sigma_{tot}$ and
$\sigma_{gas}$. For an exponential disk with central surface density
$\sigma_0$ and an exponential scale length $R_{gal}$, the total mass out
to infinity is $2\pi \sigma_0 R_{gal}^2$. Thus $\sigma_{tot}=
M_{tot}/(2\pi R_{gal}^2)$, $\sigma_{gas}=M_{gas}/(2\pi R_{gal}^2)$, and
the self-binding pressure is
$\left(GM_{gal}/R_{gal}\right)M_{gas}/\left(2\pi R_{gal}^3\right)$. This
is $(5/3)\rho_{gas}v_{esc}^2$ for $\rho_{gas}=3M_{gas}/\left(4\pi
R_{gal}^3\right)$ as above. Thus the stripping condition given by equation
(\ref{eq:strip}) with an effective galactic gas density crudely derived
for a sphere the size of the exponential scale length is the same to
within 60\% as Gunn \& Gott's condition for a disk with the gas
distributed exponentially.

In either case, the results suggest that the ram pressure from the
motion of VCC 882 through M86 was not overwhelming compared to the
internal gravitational energy density of the former interstellar medium
in that galaxy. If it were, it would probably have 
led to a more rapid loss
of gas from the dwarf, in one or two big clumps
(as for UGC 7636). Instead, the
stripping was apparently mild, and the morphology of the stripped gas
more filamentary, like a long trail of debris removed somewhat steadily.

In fact, this morphological dependence on stripping strength is in
agreement with general theory. Stevens, Acreman, \& Ponman (1999) found that
long gas trails like the one we observe here result if the stripping
pressure is modest rather than overwhelming, and if the galactic stars
continuously replenish the interstellar medium at normal rates through
winds and supernovae. Earlier calculations by Lea \& De Young (1976) got
a trail for this case too; they explained it as the result of expansion
of galactic gas into the low pressure zone downstream in the trail.
Takeda, Nulsen \& Fabian (1984) found a rapid initial loss of gas for
their models, but this was followed at long times by a trail of more
steady ablation consisting of gas from the stellar replenishment.
Balsara, Livio \& O'Dea (1994) got a trail in the mass-replenishment
case too, but found also that stripping occurred in bursts. 
For steady state models, trails occur when there is a low rate of gas
replenishment and a 
high ram pressure (Portnoy, Pistinner, \& Shaviv 1993).

The VCC 882 case may have had stripping bursts too, in addition to a
more steady gas loss that formed the overall trail. Figure 2 suggests
that the trail is not perfectly uniform, but has three prominent
condensations. These are approximately equally spaced, and consistent
with the suggestion by Balsara, Livio \& O'Dea (1994) and Stevens,
Acreman \& Ponman (1999) that the bursts of mass loss would be periodic.
The timescale for these bursts should depend on the turbulent or sound
crossing time inside the dwarf galaxy, out to the shocks at the
interacting surfaces (cf. Lea \& De Young 1976), because internal
adjustments and pulsations in the interstellar medium of the dwarf have
this characteristic scale. The positions of these
shocks depend on the flow speed, so the oscillation timescales should also
scale with the flow crossing time outside the galaxy, which is
$2R_{replenish}/v_{rel}$. Here $R_{replenish}$ is the mass-replenishment
half radius defined by Stevens et al. (1999). Specific timescales given
in these previous models range from 3 to 5 $\times10^7$ years, which was
several flow-crossing times (see also Abadi, Moore, \& Bower 1999). 
Here the flow-crossing time is much
shorter, $2R_{gal}/v_{rel}\sim 0.5$ My, so several flow times is perhaps
slightly over 1 My. Considering that the whole trail may be only 15 My
old (uncertain because of projection effects), this oscillation
timescale is short enough to be consistent with the occurrence of three
condensations along the trail behind VCC 882. 

Other aspects of the trail morphology might result from variable
pressures in the M86 halo, i.e., increasing with depth, or from
Kelvin-Helmholtz instabilities along the trail of gas. The wiggles in
Feature A cannot result from the rotation of VCC 882 because the
rotation period at one exponential scale length is $\sim50$ My, 
which is longer than the expected trail formation time.

The long term evolution of the trail is presumably dominated by
evaporation and mixing with the hot halo of M86. The evaporation
time given by Veilleux et al. (1999) provides a useful
guide: \begin{equation} t_{evap}=1000 n_{trail}R^2_{trail,pc}
\left(T_{M86}/10^7\;K\right)^{-5/2}\left(\ln\Lambda/30\right)
\;yr; \end{equation} $\Lambda$ is the Coulomb logarithm. The trail
density, $n_{trail}$ is unknown, but we can use a column density of
$1.9\times10^{20}$ cm$^{-2}$ from the estimated extinction of
A$_V=0.1$ mag to give $n_{trail}R_{trail,pc}=30$ (half the total because $R$
is a radius rather than the full depth of the trail).  For $R_{trail,pc}$
comparable to the observed half-width of $\sim250$ pc, the evaporation time
is  $\sim7.5$ My, which is half the estimated trail age.  Thus the
trail length could be determined in part by evaporation.

It is conceivable that VCC 882 is bound to M86 and produces a continuous
trail from replenished gas that always evaporates by the time it reaches
$10-20$ My downstream. This can be checked from the orbit speed in
the potential of M86. If we assume that the two galaxies are currently
separated by their projected distance of $D=86.9^{\prime\prime}= 7.7$ kpc,
and that VCC 882 is in a parabolic orbit with a speed of $v_{rel}=1328$
km s$^{-1}$ (ignoring projection effects), then the mass of M86 becomes
$0.5Dv_{rel}^2/G\sim 1.6\times10^{12}$ M$_\odot$. The log of the blue
luminosity of M86 is 10.65 (Tully 1988), which then gives an M/L ratio
of 35 M$_\odot$/L$_\odot$. Although the average M/L for elliptical
galaxies is 5 to 10, the M/L ratio for M86 was calculated to be 63 based
on mass estimates from the X-ray temperatures and densities (Forman
et al. 1985). Our value is consistent with their result, considering
the uncertainties here. This suggests that much of the apparent speed
of VCC 822 could be from its motion inside M86. 
The visible trail could be only the non-evaporated remnant of a longer
trail that has periodically removed fresh gas from VCC 882 as it plunged
into the denser regions of M86.

This situation is similar to the model proposed by Takeda et al. (1984)
for elliptical galaxies in clusters, applied here on a smaller scale.
However, for VCC 882, the gas mass in the trail is probably too large
to be entirely the result of stellar wind accumulation in an orbit time
around M86. For the replenishment rate of $dM/dt=1.6\times10^{-11}M_{gal}$
yr$^{-1}$ used by Balsara et al.  (1994), the time for $1.4\times10^8$
M$_\odot$ of gas to accumulate inside an $8\times10^8$ M$_\odot$
galaxy is $10^{10}$ years. This would have to be the orbit time if the
stripping process were repetitive, but it is such a long time 
compared to the Hubble time that VCC
882 would have to be on its first orbit. If the gas accumulation time
is not shorter than this, then the trail behind VCC 882 is probably a
first stripping event.  Even longer replenishment times were assumed
for elliptical galaxies by Gaetz, Salpeter \& Shaviv (1987).

The distinction between these two models of first-time and repetitive
stripping is relevant to the problem of the origin of nucleated dwarf
ellipticals. If VCC 882 is now experiencing its first stripping event,
then it is just now becoming a gas-poor elliptical galaxy and its
formation mechanism would have been one involving stripping from a
gas-rich dwarf irregular (Lin \& Faber 1983), or preferably a BCD,
considering its compact bright nucleus (Davidge 1989; Sandage \&
Hoffman 1991; Vader \& Chaboyer 1994).  If the gas accumulation rate is
faster than the Balsara et al.  estimate by a factor of 10, then VCC 882
could have formed as an elliptical long ago by other mechanisms
and now be shedding only the gas mass that comes from evolved stars.
The concentration of dE,N galaxies in the center of the Virgo cluster
suggest an early formation anyway (Gallagher \& Hunter 1989), consistent
with dwarf formation models in a cold dark matter universe (Silk, Wyse,
\& Shields 1987).

The origin of dE's in Virgo is unclear although their properties are
well studied.  Nucleated dE's like VCC 882 concentrate toward the center
of Virgo, unlike the non-nucleated dE's and the dI's (van den Bergh 1986;
Ferguson \& Sandage 1989).  This suggests either an early origin for
the dE's or stripping from a previous gas-rich state.  The centralized
dE,N's are also brighter and larger than the peripheral dE's (Ichikawa et
al. 1988), so stripping from a common ancestor will not produce both types
of dE's.  The nucleated dE's like VCC 882 have more globular clusters
per unit luminosity than non-nucleated dE's (Harris 1991), and this also
suggests a primordial origin for the dE,N's, unless the globulars were
made during the stripping event.  The exponential light profile in VCC
882 is typical for dE's, which are like dI's in this respect (Faber \&
Lin 1983; Caldwell 1983; Binggeli, Sandage \& Tarenghi 1984), but the
dE's have more metals than the dI's (Thuan 1985) so either the dE's
had a distinct origin or early stripping events triggered substantial
amounts of star formation.  No star formation is seen in VCC 882,
however; there is 
only a slight B-I color gradient along the minor axis (Fig. 5).
dE's are also distinct from E's and globular clusters on correlation
diagrams considering galactic structure parameters (Binggeli, Sandage \&
Tarenghi 1984; Kormendy 1985), but not considering color (Caldwell 1983;
Bothun \& Caldwell 1984; Wirth \& Gallagher 1984; Zinnecker et al. 1985),

These properties of dE,N galaxies suggest that most of them acquired their
morphologies early on, possibly passing through an early BCD phase to get
the bright nucleus and/or globular clusters through intense
star formation.  Indeed, dE,N nuclei resemble bright globulars
(Phillips et al. 1996).  But in this case the presence of what
appears to be a trail following VCC 882 is an anomaly. This galaxy
could not have been a BCD only $\sim15$ My ago, which is the stripping
time considering the trail length and relative velocity of M86,
because that would make VCC 882 too blue now.  It might have had a
mixed morphology before, like ESO 359-G29 (Sandage \& Fomalont 1993)
or several other dwarf galaxies  (Vigroux et al. 1986; Davidge 1989;
Sandage \& Hoffman 1991; Vader \& Chaboyer 1994).  Or it could have
an anomalously high gas replenishment rate, which would make the trail
come from stripped stellar ejecta in a more continuous flow.  The lack
of obvious blue stars or young stellar clumps in what was supposedly a
gas-rich system only several tens of millions of years ago is difficult
to understand. Perhaps the trail did not come from VCC 882 at all.

Observations of the metallicity, dust-to-gas ratio, grain properties,
and velocities in the trail would be useful, as would more detailed
studies of the current stellar populations in VCC 882. High resolution
X-ray observations of the region around and behind VCC 882 would be
interesting too, to distinguish the trail from the plume and other X-ray
emitting gas connected with M86.

\section{Conclusions} 

Deep B, V, and I images of the southeast core region of the Virgo
cluster reveal a dust trail in the central region of M86, spanning 28
kpc with a thickness of $\sim500$ pc. The dust appears to connect to the
nucleated dwarf galaxy VCC 882 located just north of M86. We estimate
that the dust has a V-band extinction of 0.1 mag, with a total mass of
$10^8$ M$_\odot$. This dust and the associated gas may have been
ram-pressure stripped from VCC 882 over the last 10--20 My as the dwarf
galaxy passed through the hot X-ray emitting gas of M86. The mass of gas
is somewhat consistent with its former presence inside VCC 882, and the
morphology of a trail is consistent with theoretical predictions for
relatively weak ram pressures and periodic outflows. The similarity
between the evaporation time of the trail and the dwarf orbit time
suggest that evaporation is important. The length of the trail could
even be determined by steady evaporation at the back end. 

A problem with this interpretation is that VCC 882 is a classical
nucleated dwarf.  It seems to have no recent star formation as if
it just had a gas disk, and the gas mass estimate is much too high
for such a morphology. If the trail really did come from VCC 882,
then it would have had a mixed morphology before the stripping,
like several other rare cases that have been studied elsewhere. 
More observations will be required to determine the origin and
history of this gas. 

We gratefully acknowledge summer student support from the Wm. Keck
Foundation to the Keck Northeast Astronomy Consortium, and thank Vassar
College for a research publication grant.

\newpage
\begin{figure}
\vspace{4.0in}
\includegraphics{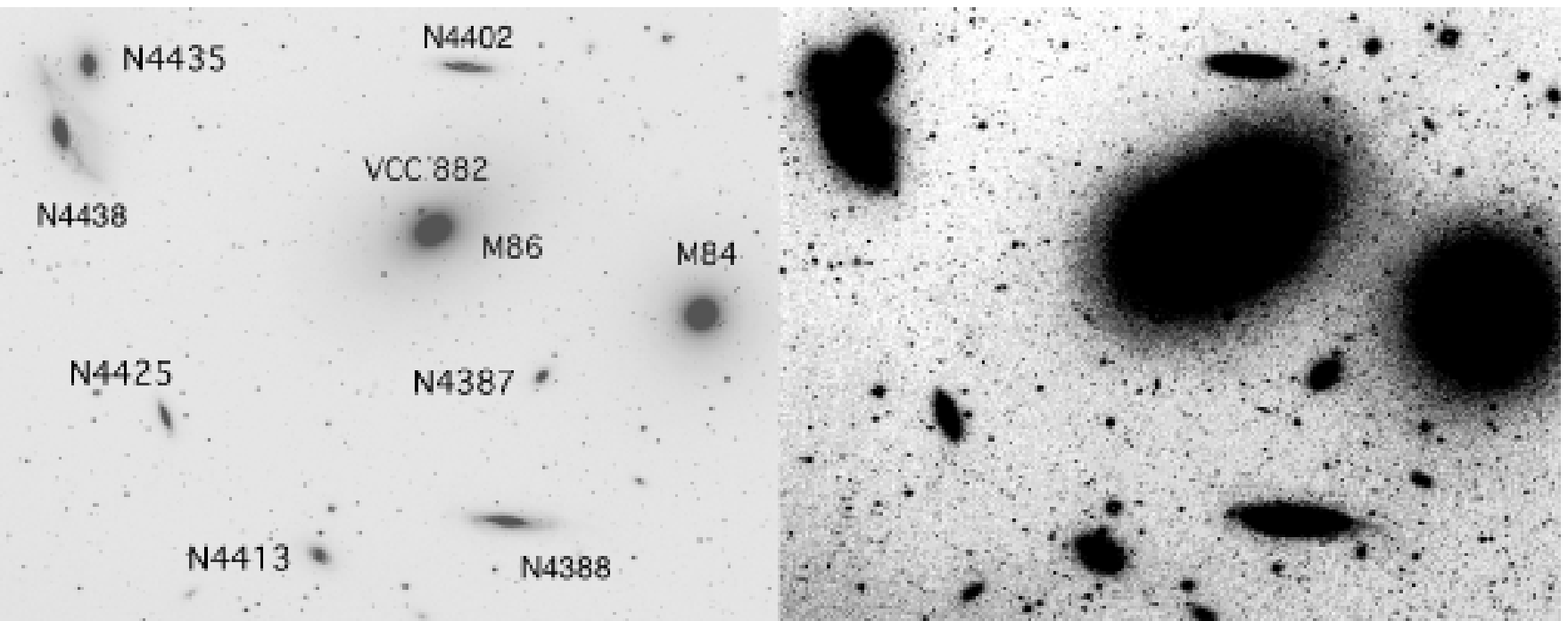}
\caption{B band images of the central region of the Virgo cluster,
with two contrast levels. 
{\it The image bpi has been reduced here by 4 to save space for astro-ph.}
}
\end{figure}

\newpage
\begin{figure}
\vspace{6.0in}
\includegraphics{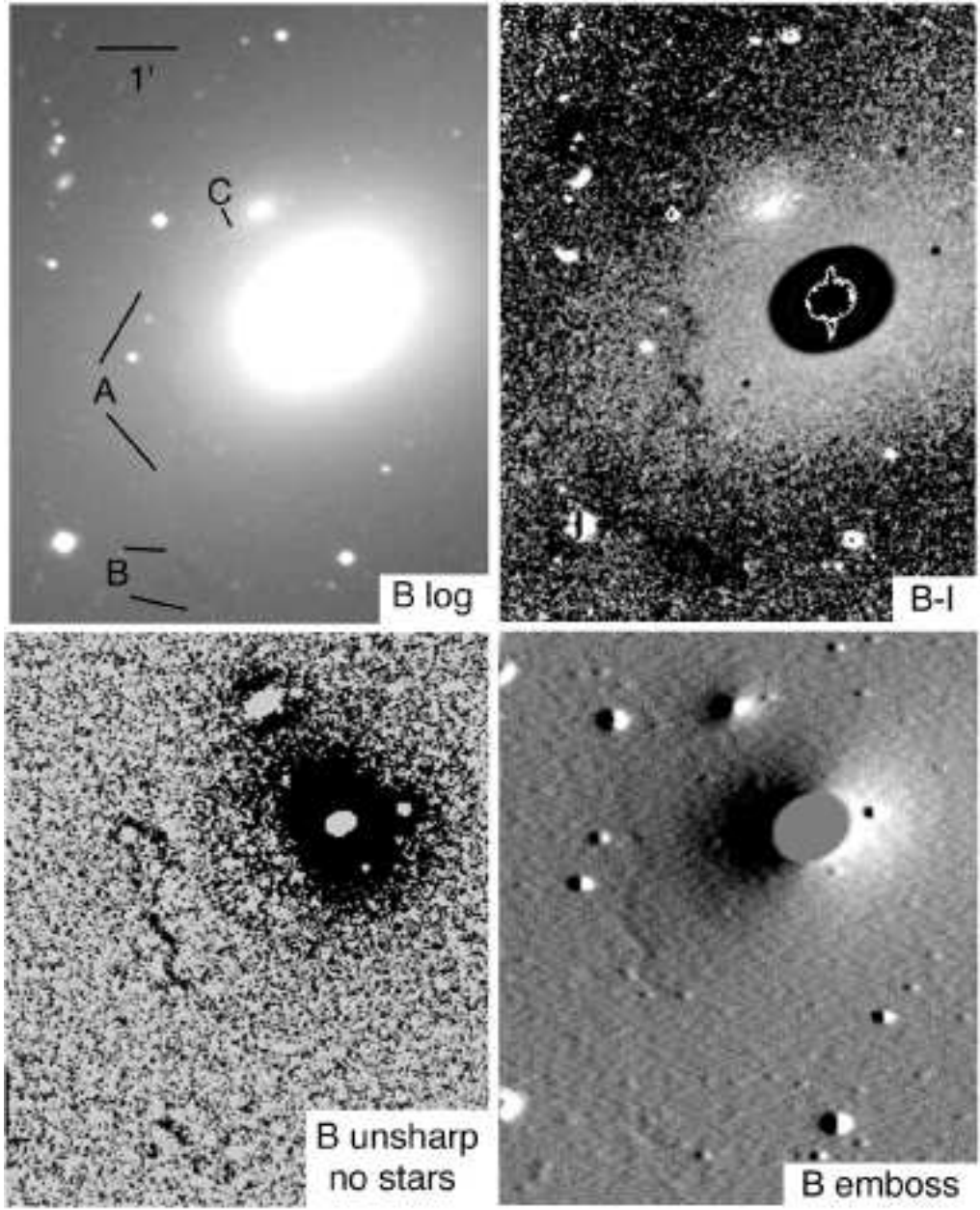}
\caption{Enhanced images of M86 with VCC 882 to the northeast.
Top left: logarithmic intensity in the B band; top right: B-I color
image; bottom left: unsharp-mask restoration of small-scale structure
in the B band; bottom right: embossed representation of the B band image.
{\it The image bpi has been reduced here by 4 to save space for astro-ph.}
}
\end{figure}

\newpage
\begin{figure}
\vspace{6.0in}
\includegraphics{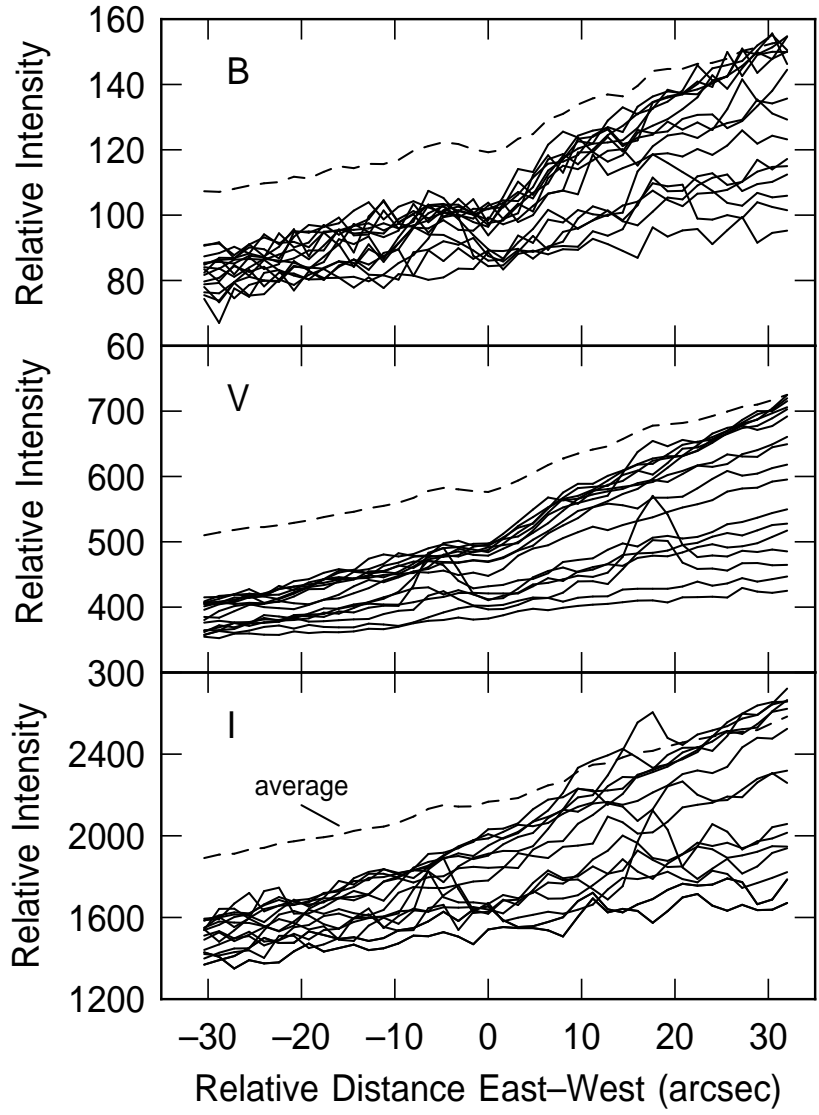}
\caption{Intensity strips across Feature A in three passbands, 
B at the top, V in the middle and I at the bottom. The dust filament
is approximately at the relative distance 0.} 
\end{figure}

\newpage
\begin{figure}
\vspace{5.0in}
\includegraphics{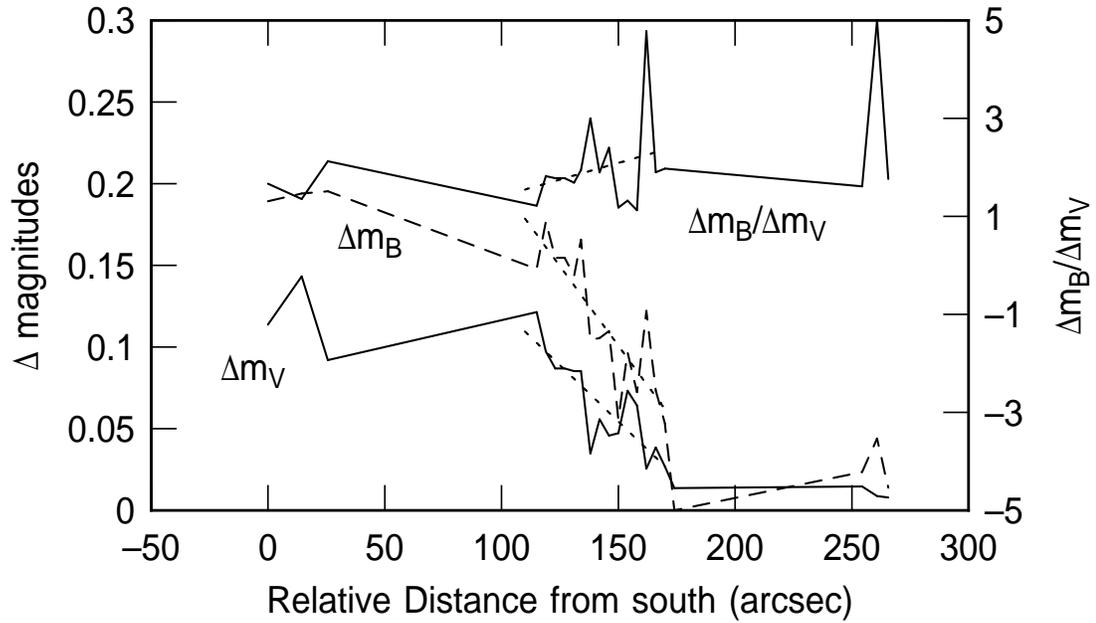}
\caption{Magnitude differences between the trail and the surrounding
regions are plotted as a function of distance along the trail, from 
south to north. 
The three plotted points on the left are Feature B, the cluster
of points in the middle is Feature A, and the three points on the
right are Feature C. }
\end{figure}

\newpage
\begin{figure}
\vspace{5.0in}
\includegraphics{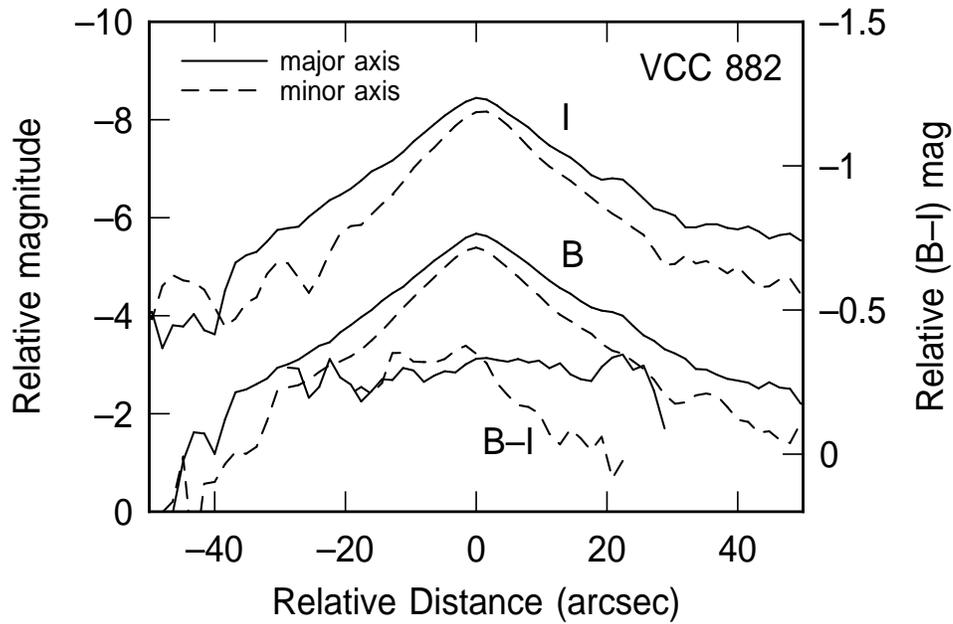}
\caption{B and I-band profiles of relative intensity
along the major and
minor axes of VCC 882.  The shift between the profiles
is arbitrary.  The major axis profile is an average over 11 adjacent
profiles spaced by one pixel in the minor axis direction, 
and the minor axis profile is an average over 21 profiles
spaced by one pixel in the major axis direction. 
The 
underlying light from M86 was remove.
The profiles are 
exponential 
with a major axis scale length of 3.8$^{\prime\prime} = 340$ pc. The
(B-I) color profiles along the major and minor axes are also
plotted at the bottom of the diagram, using the scale on the right. 
}
\end{figure}

\newpage
\begin{figure}
\vspace{5.0in}
\includegraphics{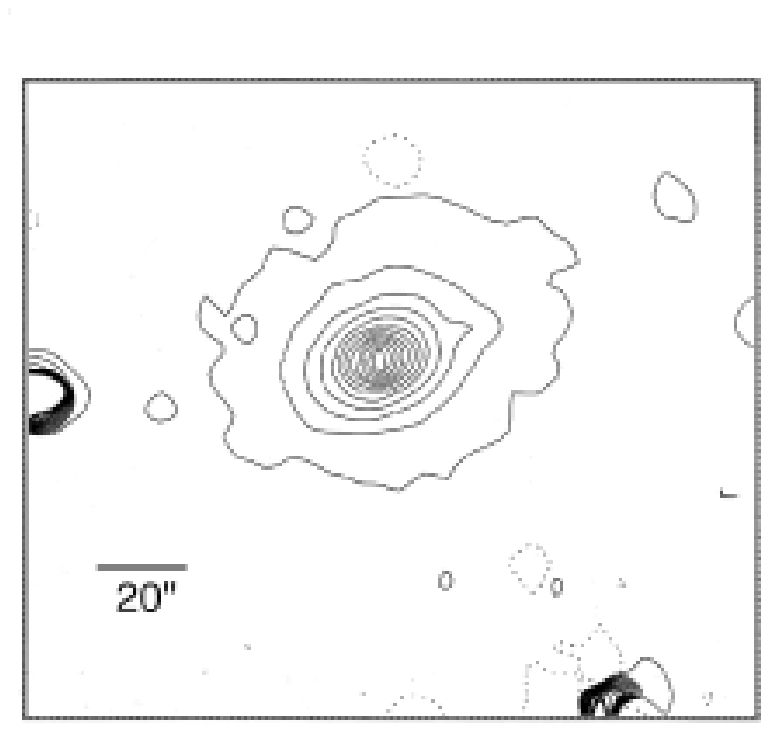}
\caption{ Isophotal contours of VCC 882 in B band. No distortion 
from a stellar tidal tail is evident.
{\it The image bpi has been reduced here by 4 to save space for astro-ph.}
}
\end{figure}

\newpage
\begin{figure}
\vspace{5.0in}
\includegraphics{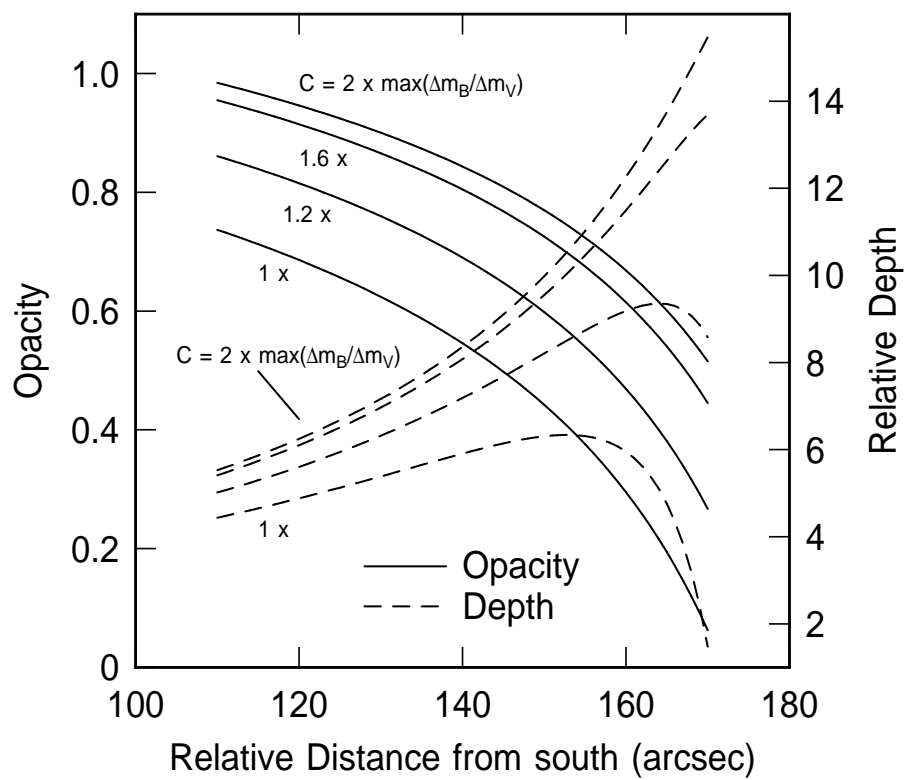}
\caption{Radiative transfer model fits for the
opacities (solid lines) and relative depths (dashed lines) of the dust
trail in the stellar halo of M86. The linear fits to the
magnitude differences in Fig. 4 are used. The parameter $C$ is the
intrinsic ratio of B to V extinction, equal to 1.3 for a
Whitford law. Here is it much higher, taken to be the indicated factors
times the maximum observed value of the B-to-I ratio, which is
2.36. The large value of $C$ suggests the dust grains in the trail
are smaller than in the local interstellar medium. }
\end{figure}
\end{document}